\documentclass[prl,twocolumn,showpacs,floats]{revtex4}

\usepackage{graphicx}
\usepackage{booktabs}
\usepackage{tabularx}

\usepackage{CJK}

\begin{document}
\begin{CJK*}[pmC]{GB}{song}

\title{Growth, Characterization and Fermi Surface of Heavy Fermion CeCoIn$_5$ Superconductor}
\author{JIA Xiao-Wen$^{1}$, LIU Yan$^{1}$, YU Li$^{1}$, HE Jun-Feng$^{1}$, ZHAO Lin$^{1}$, ZHANG Wen-Tao$^{1}$, LIU Hai-Yun$^{1}$, LIU Gong-Dong$^{1}$, HE Shao-Long$^{1}$, ZHANG Jun$^{1}$, LU Wei$^{1}$, WU Yue$^{1}$, DONG Xiao-Li$^{1}$, SUN Li-Ling$^{1}$, WANG Gui-Ling$^2$, ZHU Yong$^2$, WANG Xiao-Yang$^{2}$, PENG Qin-Jun$^{2}$, WANG Zhi-Min$^{2}$, ZHANG Shen-Jin$^{2}$, YANG Feng$^{2}$, XU Zu-Yan$^{2}$, CHEN Chuang-Tian$^{2}$ and ZHOU Xing-Jiang$^{1,*}$}

\affiliation{
\\$^{1}$National Lab for Superconductivity, Beijing National Laboratory for Condensed
Matter Physics, Institute of Physics, Chinese Academy of Sciences,
Beijing 100190, China
\\$^{2}$Technical Institute of Physics and Chemistry, Chinese Academy of Sciences, Beijing 100190, China
}
\date{January 18, 2011}
%
%

\begin{abstract}

High quality single crystals of heavy Fermion CeCoIn$_5$ superconductor have been grown by flux method with a typical size of (1$\sim$2)mm $\times$ (1$\sim$2)mm $\times$ $\sim$0.1 mm. The single crystals are characterized by structural analysis from X-ray diffraction and Laue diffraction, as well as compositional analysis. Magnetic and electrical measurements on the single crystals show a sharp superconducting transition with a transition temperature at T$_{c(onset)}$ $\sim$ 2.3 K and a transition width of $\sim$0.15 K.   The resistivity of the CeCoIn$_5$ crystal exhibits a hump at $\sim$45 K  which is typical of a heavy Fermion system.  High resolution angle-resolved photoemission spectroscopy (ARPES) measurements of CeCoIn$_5$ reveal clear Fermi surface sheets that are consistent with the band structure calculations when assuming itinerant Ce 4f electrons at low temperature.  This work provides important information on the electronic structure of heavy Fermion CeCoIn$_5$ superconductor.  It also lays a foundation for further studies on the physical properties and superconducting mechanism of the heavy Fermion superconductors.

\end{abstract}


\maketitle



Heavy Fermion materials are characterized by a significant enhancement of the electronic specific heat at low temperature, corresponding to a dramatic enhancement of the effective mass of charge carriers\cite{zfisk,GRStewart}. They usually involve rare-earth or actinide elements such as Uranium (U) or Cerium (Ce) which have 4f or 5f electrons\cite{fisher,hrott,geibel1,geibel2,steglich,petrovik}. Heavy fermion systems have attracted much attention because they are an ideal platform to study rich physics like non-Fermi liquid behaviors, quantum critical behaviors, unconventional superconductivity, the interplay of superconductivity and magnetism and etc.\cite{zfisk,GRStewart}. In the Landau Fermi liquid picture,  electrons and all direct or indirect interactions between them are treated in terms of a ``quasi-particle" without effective mutual interaction. In heavy Fermion materials, this treatment is no longer valid as shown in experimental results that no constant specific heat coefficient $\gamma$ and magnetic susceptibility are observed at low temperature\cite{seaman}. Theoretically, the origin of the non-Fermi-liquid behavior in heavy Fermion materials remains under debate that involves various models proposed\cite{amoto,yyfeng,yonuki}. One is the screening of local moments by conducting electrons, i.e., the Kondo model, where the competition between Ruderman-Kittel-Kasuya-Yosida (RKKY) interaction and Kondo effect dictates the low temperature behavior. The other is the critical point transition under external field (magnetic field or pressure) or doping. The proximity to a magnetic order in the phase diagram indicates that a magnetic fluctuation at \textit{T}=0 may contribute to the non-Fermi-liquid behavior at high temperature. The third possibility is the disorder effect: the Kondo temperature can be reduced by disorder which may induce non-Fermi-liquid behavior.

It is generally believed that magnetism is detrimental to superconductivity because, according to the BCS theory of superconductivity\cite{BCS}, the local magnetic moments may break the spin-singlet state of the Cooper pairs.  Therefore, the discovery of superconductivity in heavy Fermion materials is surprising and particularly interesting\cite{steglich}. It provides an ideal system to study the relation and interplay between magnetism and superconductivity.  It also provides opportunity for the realization of unconventional superconductivity where the pairing of Cooper pairs is not mediated by phonons as in the BCS theory but possibly by electron-electron interaction\cite{amoto,steglich}.

Until now, a large number of heavy Fermion superconductors have been discovered, such as U-based UPt$_3$,UBe$_{13}$, UPd$_2$Al$_3$, UNi$_2$Al$_3$, and Ce-based CeCu$_2$Si$_2$, and so on\cite{fisher,hrott,geibel1,geibel2,steglich,petrovik}. Among them, the recently discovered CeMIn$_5$ (M=Co,Rh,Ir), the so-called 115 heavy Fermion family, is especially interesting\cite{JDThomson}.  CeCoIn$_5$  and CeIrIn$_5$ undergo a superconducting transition at 2.3 K and 0.4 K, respectively, while CeRhIn$_5$ orders antiferromagnetically with T$_N$=3.8 K at ambient pressure and becomes superconducting with a T$_c$=2.1 K under high pressure\cite{petrovik,JDThomson}.   CeCoIn$_5$ stands out to be particularly interesting because of a number of reasons. First, it has a high superconducting transition temperature (T$_c$=2.3 K) among all known heavy Fermion superconductors\cite{petrovik,tpark,egmoshopolow,hhegger,Akoitzsch1}. In fact, its T$_c$ is second only to heavy Fermion PuCoGa$_5$ with a T$_c$=18 K\cite{njcurro}. However, the radioactivity and rareness of Plutonium (Pu) render PuCoGa$_5$ difficult to be well-studied. Therefore, the relatively high T$_c$ of CeCoIn$_5$ and its easy availability make it possible to carry out critical experiments like angle-resolved photoemission spectroscopy (ARPES) to probe its superconducting state.  Second, transport and other measurements indicate that superconductivity in CeCoIn$_5$ is unconventional and favors a d$_{x2-y2}$ gap symmetry\cite{izawa} or a d$_{xy}$ gap symmetry\cite{HAoki}. Third, the critical point transition was reported in CeCoIn$_5$\cite{biachi}. Fourth, in the superconducting state, it was reported that an FLLO state may be realized in CeCoIn$_5$\cite{biachifllo}.

The key to understand the non-Fermi-liquid behaviors, the unconventional superconductivity, and other significant issues in the heavy Fermion materials is to have a complete understanding of  their underlying electronic structure.  Specifically, how heavy Fermion state is realized by the interaction of f electrons with the conduction carriers to become either localized or itinerant.  ARPES is a powerful tool to directly probe the electronic structure of materials\cite{damaceli}.  However, so far only few ARPES work have been reported for CeCoIn$_5$\cite{Akoitzsch1,Akoitzsch2} and many issues remain to be addressed. In this paper, we report successful growth of high quality single crystals of CeCoIn$_5$ superconductor and high resolution ARPES measurements on its Fermi surface. Clear Fermi surface sheets of CeCoIn$_5$ are revealed which are consistent with the band structure calculations assuming Ce 4f electrons to be itinerant at low temperature. This work provides important information on the electronic structure of CeCoIn$_5$ superconductor. It also lays a foundation for further ARPES studies on its physical properties and unconventional superconducting mechanism.

\begin{figure}[tbp]
\begin{center}
\includegraphics[width=0.95\columnwidth,angle=0]{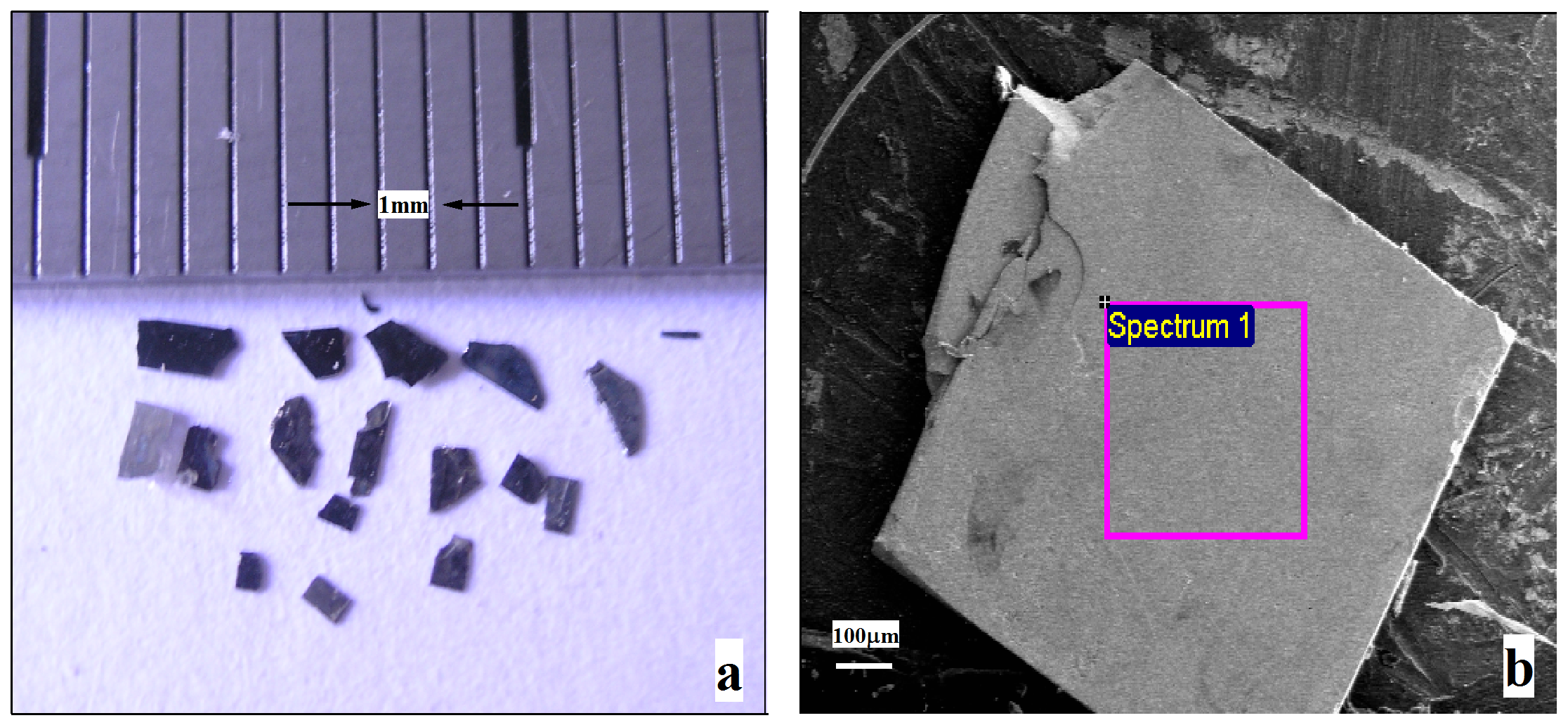}
\end{center}
\caption{(a). Photographs of the as-grown CeCoIn$_5$ single crystals. The crystal size is typically (1$\sim$2) mm with a thickness of $\sim$ 0.1 mm. (b). The surface morphology of  CeCoIn$_5$ single crystal observed by SEM.  The square represents the area where the composition is measured by EDAX. }
\end{figure}

High quality single crystals of CeCoIn$_5$ were grown by using the flux method\cite{petrovik}. Ce ingot, Co powder and In spheres were mixed according to an atomic ratio of 3:3:94; the excess In was used as flux. Before the mixing, the Ce ingot was  polished carefully to remove the oxidized surface and cut into small pieces. The  starting materials were mixed thoroughly and put into a cylinder-shaped crucible, some quartz wool were then placed in the top of the crucible but did not touch the mixed starting materials. Then the crucible was encapsulated in a highly evacuated quartz tube. We found that the growth of CeCoIn$_5$  single crystals depends sensitively on growth temperature and the cooling rate. We tried different conditions and used the following procedure to obtain large size and high quality CeCoIn$_5$ crystals: (1) quickly heat up the mixed starting materials to 1150 $^{\circ}$C in 3 hours; (2) stay for a few minutes at 1150 $^{\circ}$C; (3) rapidly cool down to 750 $^{\circ}$C in 4 hours; (4) slowly cool down to 450 $^{\circ}$C in 100 hours. After this, the crucible was quickly moved to a centrifuge to separate the single crystals from the excess In which passed through the quartz wool.

\begin{figure}[b]
\begin{center}
\includegraphics[width=0.95\columnwidth,angle=0]{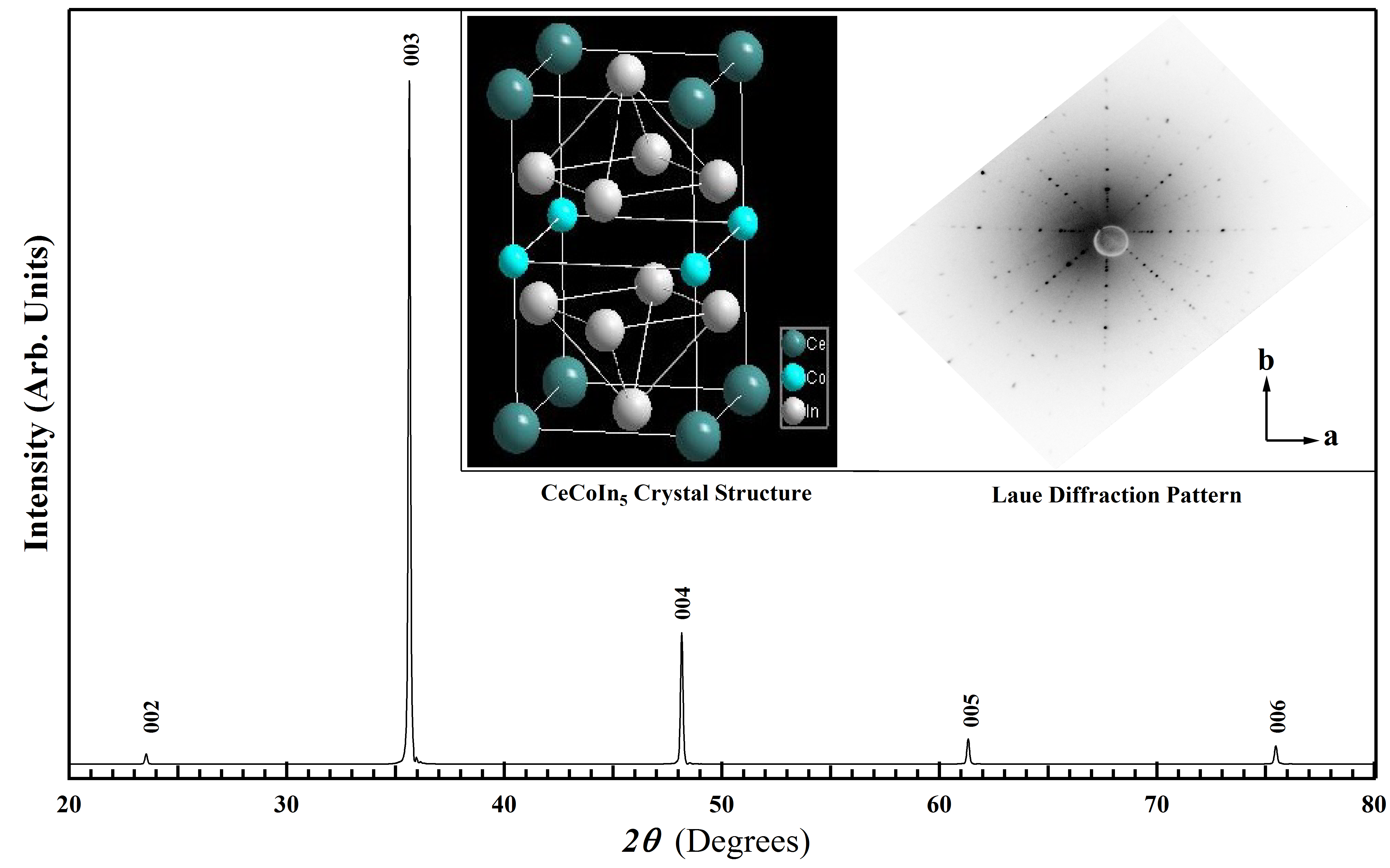}
\end{center}
\caption{X-Ray diffraction pattern of the CeCoIn$_5$ single crystal. All the peaks can be indexed by (0,0,h) of CeCoIn$_5$ and the fitted c parameter is 7.5515(1){\AA}. The left side of the upper-right inset shows the crystal structure of CeCoIn$_5$, while the right side is a Laue diffraction pattern with the crystal orientation indicated by black arrows. Note that a and b are equivalent due to P4/mmm space group of CeCoIn$_5$.}
\end{figure}

\begin{table}[htbp]
\caption{SEM$\slash$EDAX analysis result on the composition of the as-grown single crystals.\label{tab:table1}}
\begin{tabular*}{0.45\textwidth}{lccc}
\hline
Element & ~~~~~~~~~~~~~~~~~~Weight\% & ~~~~~~~~~~~~~~~~~~Atomic\% \\
\hline
Ce & ~~~~~~~~~~~~~~~~~~18.34 & ~~~~~~~~~~~~~~~~~~10.69 \\
Co & ~~~~~~~~~~~~~~~~~~7.39 & ~~~~~~~~~~~~~~~~~~10.25  \\
In & ~~~~~~~~~~~~~~~~~~68.00 & ~~~~~~~~~~~~~~~~~~48.40  \\
\hline
\end{tabular*}
\end{table}

Plate-like CeCoIn$_5$ single crystals with shiny surface are obtained, as shown in Fig. 1a, with a typical size of (1$\sim$2) mm $\times$ (1$\sim$2) mm $\times$ $\sim$0.1 mm.   The surface morphology and the chemical composition were characterized by scanning electron microscope (SEM) equipped with the energy dispersive X-ray spectroscopy (EDAX) (Fig. 1b).  The obtained CeCoIn$_5$ crystals show smooth surface (Fig. 1b). The average composition of the crystals is consistent with the chemical composition of Ce:Co:In=1:1:5 within the error bar of the EDAX analysis (Table 1).

Fig. 2 shows the X-ray Diffraction(XRD) pattern of a CeCoIn$_5$ single crystal.  All the diffraction peaks can be indexed by (0,0,h) with h being an integer; the fitted c parameter is 7.5515(1){\AA} which is consistent with result reported before\cite{wkpark,PSnormile}. The left side of the upper-right inset shows the crystal structure of CeCoIn$_5$ which consists of alternate stacking of (CeIn$_3$) unit and (CoIn$_2$) unit along the c-axis. In fact, CeCoIn$_5$ corresponds to the m=1 and n=1 member in the homologous series of Ce$_n$Co$_m$In$_{3n+2m}$ which consists of sequential stacking of n layers of  (CeIn$_3$) units and m layers of (CoIn$_2$) units along the c axis\cite{JDThomson}. The right side of the upper-right inset shows the Laue diffraction pattern of a CeCoIn$_5$ crystal. The orientation of the crystal is marked in the inset of Fig. 2. The sharp peaks in the XRD pattern and the sharp spots in the Laue diffraction pattern indicate high crystallinity of the as-grown single crystals.

The physical properties of the CeCoIn$_5$ single crystals are characterized by both electrical and magnetic measurements (Fig. 3). A clear superconducting transition is observed from both the electrical and magnetic measurements with a transition temperature (onset) T$_c$ at $\sim$ 2.3 K (see the inset of Fig. 3) which is consistent with that reported before\cite{petrovik}.  The superconducting transition is very sharp with a transition width of $\sim$0.15 K (see the inset of Fig. 3) that indicates a high quality of the single crystals. The a-b plane resistance-temperature dependence of CeCoIn$_5$ exhibits a hump near 45 K: above 45 K, the resistance first decreases slowly and then increases with decreasing temperature while below 45 K the resistance decreases with decreasing temperature dramatically and eventually get into the superconducting state at about 2.3 K.  This is a typical behavior of heavy Fermion systems which can be understood in terms of competition between the RKKY interaction and the Kondo effect, as mentioned above.

\begin{figure}[tbp]
\begin{center}
\includegraphics[width=0.95\columnwidth,angle=0]{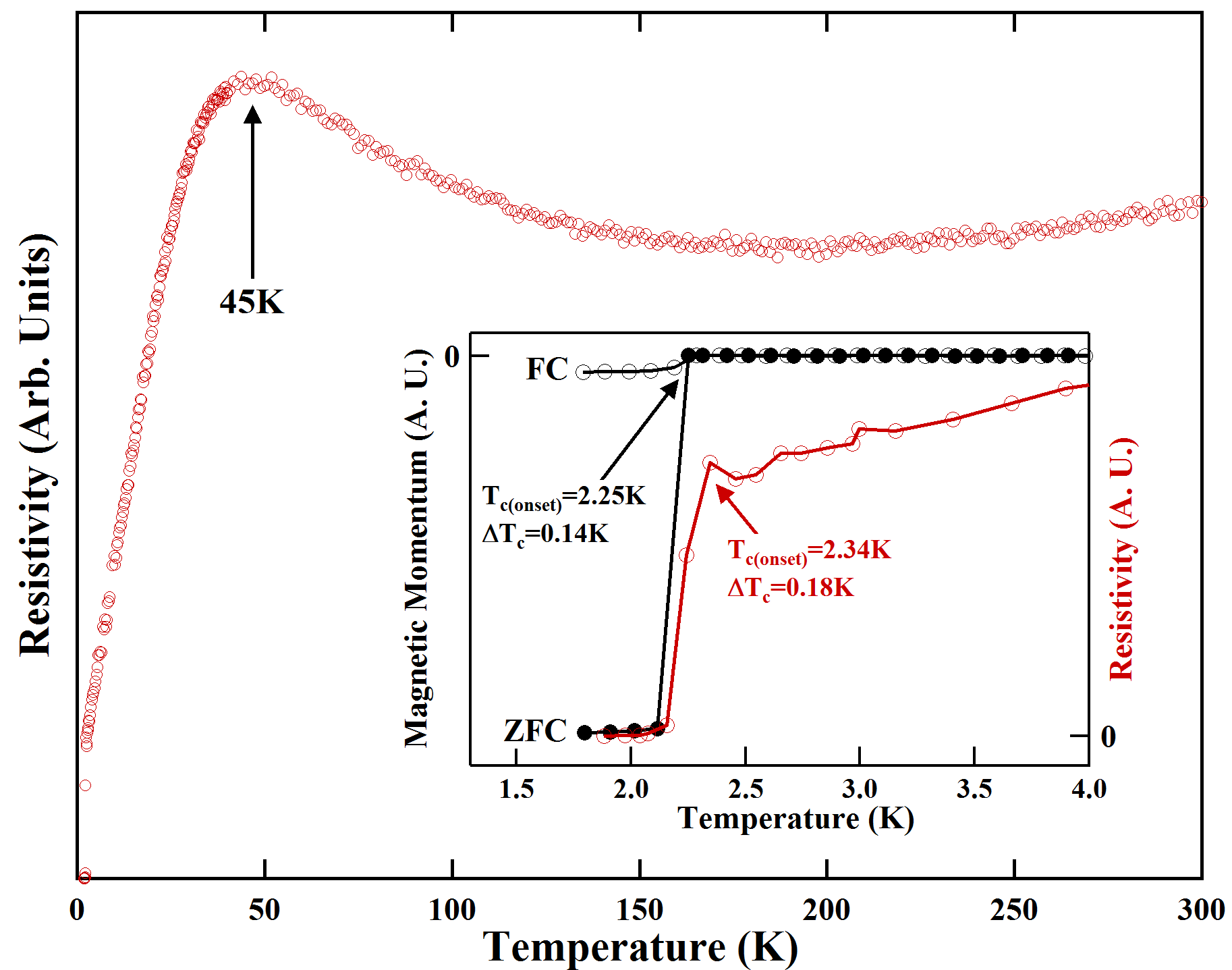}
\end{center}
\caption{Electrical and magnetic measurements of CeCoIn$_5$ single crystals. There is a clear hump near 45 K and a sharp superconducting transition at $\sim$2.3 K in the resistivity curve. The inset shows the sharp superconducting transition for both the electrical and magnetic measurements in a narrow temperature range.  The resistivity curve in the inset (violet circles) shows a clear superconducting transition at T$_{c(onset)}$=2.34 K with a transition width of 0.18 K .  In the magnetic measurement with FC (filed cooling) and ZFC (Zero filed cooling) modes shown in the inset (black solid and empty circles), the ZFC curve shows a clear superconducting transition at T$_{c(onset)}$=2.25 K with a transition width of 0.14 K.
}
\end{figure}

We carried out high resolution ARPES measurements on CeCoIn$_5$ single crystals by using our lab ARPES system equipped with Scienta R4000 analyzer and VUV5000 UV source which gives a photon energy of Helium I at h$\upsilon$= 21.218 eV\cite{gdliu}. The overall energy resolution is 20 meV and the angular resolution is $\sim$0.3 degree, corresponding to a momentum resolution of 0.009{\AA}$^{-1}$ at the photon energy of 21.218 eV. The CeCoIn$_5$ single crystals were cleaved {\it in situ} and measured in ultra-high vacuum chamber with a base pressure better than 5 $\times$ 10$^{-11}$ Torr.

Fig. 4a shows the spectral weight distribution near the Fermi level as a function of momentum measured at 25 K. This temperature is below the characteristic temperature 45 K where the resistance shows a hump (Fig. 3). The high spectral intensity contour in the mapping (Fig. 4a) represents the underlying Fermi surface.  Fig. 4c and Fig .4d show band structure measured along two typical cuts through $\Gamma$ (Cut $\#$2) and M (Cut $\#1$) points shown as red lines in Fig. 4a.  One can clearly observe a diamond-shaped hole pocket around the $\Gamma$ point, and an electron-like pocket around the M point. In addition, one can also see features in between the hole-like pocket around $\Gamma$ point and electron-like pocket near M point that forms a square-shaped high-intensity contour. Some weak feature can be seen around Y point. We note that these observations are rather different from the previous ARPES result where no hole-like pocket around the $\Gamma$ point was observed\cite{Akoitzsch1,Akoitzsch2}. The difference may be related to the  photoemission matrix element effect because they used  different photon energy (100 eV) from ours (21.2 eV)\cite{Akoitzsch1,Akoitzsch2}. It is also possible that, because of the three-dimensional character of the Fermi surface in CeCoIn$_5$, Fermi surface sheets at different k$_z$ are measured at different photon energies. The resolution of the discrepancy needs further detailed ARPES study on the photon energy dependence of the Fermi surface.

\begin{figure*}[tbp]
\begin{center}
\includegraphics[width=1.9\columnwidth,angle=0]{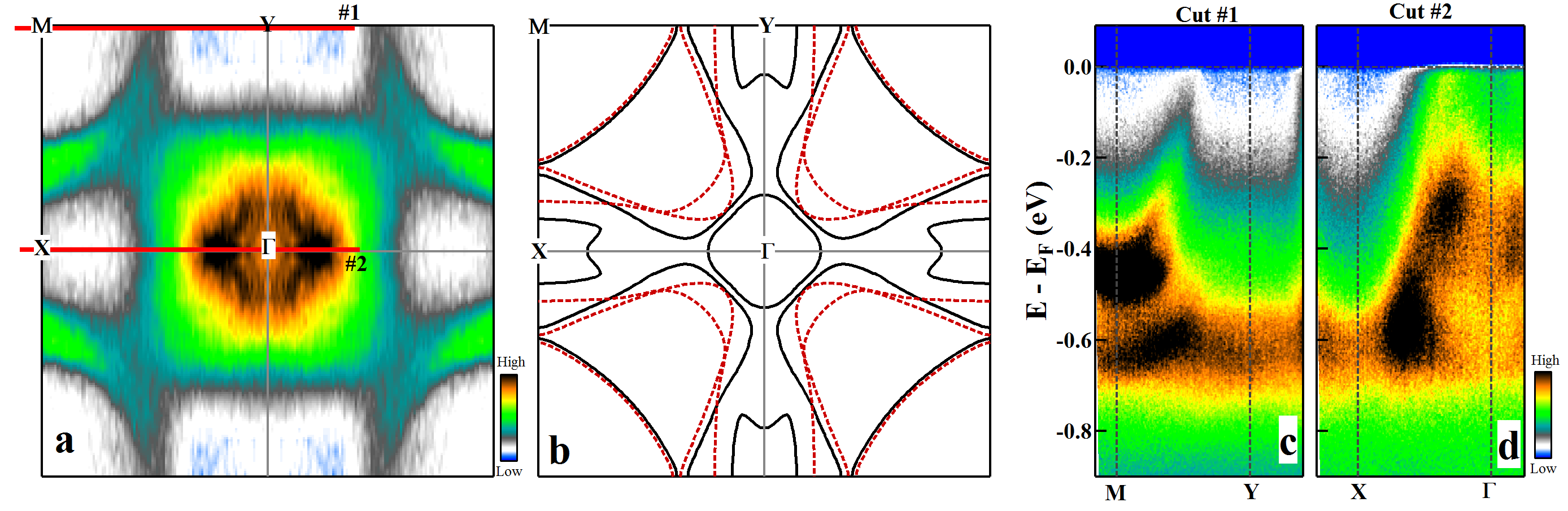}
\end{center}
\caption{Fermi surface of CeCoIn$_5$.  (a). Fermi surface mapping of CeCoIn$_5$ measured at 25 K obtained by integrating the spectral weight near the Fermi level within a small energy window.  Fermi surface sheets around $\Gamma$ point and M point are clearly seen. (b). Calculated Fermi surface topology of CeCoIn$_5$ for k$_z$=0 with Ce 4f electron assumed as being itinerant (black solid lines) and localized (violet dashed line).(c) and (d) show band structure measured along the Cuts $\#$1 and $\#$2 shown in (a) as red thick lines. Clearly the Fermi pocket around $\Gamma$ is hole-like while the one around M is electron-like.}
\end{figure*}

To understand the ARPES results, we performed band structure calculations using the Wien2K program\cite{wien2k}. Here the Ce 4f electrons are treated as itinerant or localized with spin-orbital coupling is considered. Fig. 4b shows the calculated Fermi surface both by itinerant(black solid line) and localized (violet dashed line) Ce4f model at k$_z$=0.  Apparently the calculated Fermi surface assuming Ce4f electrons as itinerant is more consistent with our ARPES measurement. The calculation based on the localized model gives no hole pocket around the $\Gamma$ point which is apparently different from our measured result. By carefully comparing our results with the itinerant model-based calculation, consistencies can be found as follows: (1). Hole-like pocket around $\Gamma$ point and electron-like pocket around M point ; (2). Weak feature at Y point; (3). Feature between $\Gamma$ and M points.  The overall consistency between the measured Fermi surface sheets and the calculated ones based on itinerant Ce4f model indicates that, in CeCoIn$_5$, the Ce 4f electrons tend to be itinerant at low temperature.  This conclusion is different from the previous ARPES results which suggested a localized 4f character in CeCoIn$_5$\cite{Akoitzsch1,Akoitzsch2}.It should be noticed that
here we omit the fact that photon energy of certain
value corresponds to a specific kz plane in the reciprocal
space. Further work needs to be carried out to
investigate the kz effect by using different photon energies.

In summary, single crystals of heavy Fermion superconductor CeCoIn$_5$  have been successfully grown by flux method. These crystals have high quality as characterized by X-ray diffraction and Laue diffraction measurements. They exhibit a sharp superconducting transition at T$_c$ $\sim$ 2.3 K with a transition width of $\sim$0.15 K. Our high resolution ARPES measurements reveal clear Fermi surface of CeCoIn$_5$.  The comparison with band structure calculations suggests that the Ce 4f electrons tend to be itinerant at low temperature.  These work provide important information on the electronic structure of CeCoIn$_5$. It also lays a foundation for further ARPES study on the superconductivity mechanism and other significant issues in heavy Fermion superconductors.

We thank Prof. G. C. Che for experimental assistance.  This work is supported by the NSFC (10734120) and the MOST of China (973 project
No: 2011CB921703).

$^{*}$Corresponding author:

XJZhou@aphy.iphy.ac.cn



\end{CJK*}
\end{document}